\documentclass[runningheads]{llncs}

\usepackage[T1]{fontenc}
\usepackage{times}
\usepackage{soul}
\usepackage{url}
\usepackage[hidelinks]{hyperref}
\usepackage[utf8]{inputenc}
\usepackage[small]{caption}
\usepackage{graphicx}
\usepackage{amsmath}
\usepackage{amssymb}
\usepackage{amsfonts}
\usepackage{booktabs}
\usepackage{algorithm}
\usepackage{algorithmic}
\usepackage[dvipsnames]{xcolor}
\usepackage{siunitx}
\usepackage{multirow}
\usepackage{varwidth}
\usepackage[symbol]{footmisc}
\newcommand{\deepwriting}[0]{\textbf{Deepwriting}}
\newcommand{\iamondb}[0]{\textbf{IAMonDB}}
\newcommand{\vnondb}[0]{\textbf{VNonDB}}
\newcommand{\maths}[0]{\textbf{Math}}
\newcommand{\tacotron}[0]{\textbf{Tacotron}}
\newcommand{\transformer}[0]{\textbf{Transformer}}
\newcommand{\raw}[0]{\textbf{raw}}
\newcommand{\curve}[0]{\textbf{curve}}
\newcommand*\samethanks[1][\value{footnote}]{\footnotemark[#1]}

\begin{document}

\title{Sampling and Ranking for Digital Ink Generation on a tight computational budget}

\author{
Andrei Afonin\inst{1}\thanks{work done as a student researcher at Google Research, Z\"urich, Switzerland}\thanks{These authors contributed equally to this work and share first authorship} \and
Andrii Maksai\inst{2}\samethanks[2] \and
Aleksandr Timofeev\inst{1}\samethanks[1] \and
Claudiu Musat\inst{2}
}

\institute{EPFL, Lausanne, Switzerland \and
Google Research, Z\"urich, Switzerland}

\maketitle

\begin{abstract}

Digital ink (online handwriting) generation has a number of potential applications for creating user-visible content, such as handwriting autocompletion, spelling correction, and beautification. 
Writing is personal and usually the processing is done on-device. Ink generative models thus need to produce high quality content quickly, in a resource constrained environment.

In this work, we study ways to maximize the quality of the output of a trained digital ink generative model, while staying within an inference time budget. We use and compare the effect of multiple sampling and ranking techniques, in the first ablation study of its kind in the digital ink domain. 

We confirm our findings on multiple datasets - writing in English and Vietnamese, as well as mathematical formulas - using two model types and two common ink data representations. In all combinations, we report a meaningful improvement in the recognizability of the synthetic inks, in some cases more than halving the character error rate metric, and describe a way to select the optimal combination of sampling and ranking techniques for any given computational budget.
\end{abstract}

\section{Introduction}
\label{sec:introduction}

Digital ink (online handwriting) offers users of digital surfaces a way of expression similar to pen and paper.
This mode of expression is gaining popularity with the increasing adoption of styluses and digital pens for tablets. 
In its digital form, ink 
is a medium that offers rich possibilities for personalized intelligent assistance for creativity and productivity.
One direct way of offering the assistance is via ink synthesis,  enabling user-facing features such as handwriting autocompletion, spelling correction, beautification, assisted diagramming and sketching.

Making these assistance experiences convenient and comfortable requires maximizing the output quality of the models, while respecting privacy and latency constraints. The same is true of other types of generated content, but standards might be higher in the case of digital ink generation, for example:
\begin{itemize}
\item Since assistive handwriting content appears in the same space as the content generated by the user, it's vital that the generated content is readable and not look "out-of-place". The users of generative image models for content creation purposes might be more forgiving to model mistakes, because there the model assists in the creative process where the users don't necessarily know what exactly they are looking for.
\item Personalized assistive handwriting often requires the models to observe the user's handwriting and transfer that style to the generated output. Unlike other modalities, handwriting is a personally-identifiable data. Therefore, it is important for the models to run on-device, rather than server-side. 
\item Generating suggestions (for example when doing autocompletion in handwriting) requires the models to be fast enough to produce their suggestions before the user has moved on or decided to add new content themselves. When the content is produced too slowly, it gets in the way of the user's flow rather than helping. This problem is further exacerbated by the constraint that the models run on-device.
\end{itemize}

In this work, we aim, given a trained generative model of digital ink and a computation budget, to produce readable outputs as often as possible, under the assumption that the model is going to be run on-device. To achieve this goal, we consider two classes of approaches that work well together.

\textit{Sampling.} This constrained ink modelling problem resembles text and audio generation. 
Following the work that has been done there~\cite{baevski2018adaptive,Nucleus,RankGen,TorToiSe,zhang2022coder}, we first concentrate on using perturbed probability distributions for sampling from autoregressive models. This improves the quality within a single inference call, by picking a sampling technique that minimizes the number of repetitive or incoherent samples. Examples of generated digital ink can be found in Fig.~\ref{fig:errors}.

\textit{Ranking.} We additionally train ranking models to predict the recognizability of an ink. We employ these models by first generating a diverse set of candidates and then ranking them to select the best output. This improves the quality if the time budget allows for multiple inference calls. 

Our proposed ranking approach would actually work for any binary quality measure (like thresholded $L_2$ distance in the style embedding space for style transfer~\cite{chang2021style} or edit-aware Chamfer distance for spelling correction~\cite{inkorrect}), but we focus on recognizability, since likely for any application of digital ink synthesis, the output should be recognizable. 

Our contributions are as follows\footnote{A notebook accompanying this submission that can run inference on example models for each dataset, data representation, and model type, and includes test label sets, is available here: \url{https://colab.research.google.com/drive/1AkwmDOkEIkifbOYEBdcB9PrR_Ll-fcmz}}:

\begin{itemize}
    \item We use sampling and ranking techniques for digital ink generation, and perform an ablation study on the ranking model objective, training, and tuning. To our knowledge, ours is the first work on this topic in the digital ink space.
    \item We show that selecting appropriate sampling parameters improves the quality of the output significantly compared to the typically used baselines, across multiple datasets, model types, and data representations. 
    \item We show that ranking further improves the quality, and discover that depending on the computational budget, the highest quality ranking models may not lead to optimal quality. We provide practical way of selecting the ranking model.
\end{itemize}
 
\section{Related work}
\label{sec:related}

\paragraph{Errors in autoregressive generative models.} Autoregressive generative models often generate samples with artifacts~\cite{Nucleus}. Artifacts appear when the generation process gets stuck in either high- or low-probability regions of the sampling space, and results in two types of errors, overconfidence (usually manifested as repeated tokens)~\cite{mirostat} and incoherence errors, respectively. We show examples of such errors during Digital Ink generation process in Fig.~\ref{fig:errors}. This is also known as the likelihood trap~\cite{likelihoodtrap} and stems from exposure bias~\cite{he2020quantifying}, which is difference between training done with 'teacher forcing' and inference~\cite{scheduledSampling}.

\paragraph{Sampling.} One common way of finding the trade-off between overconfidence and incoherence errors, often used in Text-to-Speech (TTS) and Natural Language Processing (NLP), is sampling~\cite{mirostat}, which modifies the distribution from which the points in the autoregressive model are sampled. Sampling from original distribution is called ancestral sampling; popular sampling techniques that extend it include Top-K~\cite{Top-k} and Top-P, or nucleus~\cite{Nucleus} sampling. Originally introduced for text generation, they propose picking a word from the distribution of the top most likely next words, limited by either number (in Top-K) or cumulative probability (in Top-P). Variations of the sampling techniques above include Typical sampling~\cite{Typical}, which selects components closest to a dynamically selected probability, Mirostat sampling~\cite{mirostat}, which select K in Top-K sampling adaptively, and Beam search~\cite{BeamSearch}.

\paragraph{Ranking models.} Another way to improve the generation quality is to generate several samples and choosing the best one among them. This is frequently done in information retrieval domains such as question answering~\cite{liu2020asking}, text summarization~\cite{ravaut2022summareranker}, and code generation~\cite{zhang2022coder}. Approaches most similar to ours are the ones that use ranking models for conditional generative modeling. In~\cite{RankGen}, the ranking model is trained to predict the best text continuation, with positive samples coming from real text and negative samples coming from different parts of the text and model-generated continuations. In ~\cite{TorToiSe}, two ranking models are trained to predict the match between the generated audio and the target label, as well as between the generated audio and the source audio used for style extraction. They are combined with weights specified by the user, to rank audio generated with specific style.

\paragraph{Handwriting synthesis.}
Two of the most popular models for digital ink generation are multi-layer LSTMs with monotonic attention over the label~\cite{graves2013generating} (also known in TTS as Tacotron~\cite{wang2017tacotron}) and the encoder-decoder Transformer architecture~\cite{vaswani2017attention}. Other architectures include VRNN~\cite{VRNN} used in ~\cite{aksan2018deepwriting}, Neural ODEs~\cite{das2021sketchode}, and Diffusion models~\cite{Diffusion}.

These architectures underpin applications such as sketch generation~\cite{ha2017neural} and completion~\cite{ribeiro2020sketchformer}, style transfer~\cite{kotani2020generating}, beautification~\cite{aksan2018deepwriting}, spelling correction~\cite{inkorrect}, and assisted diagramming~\cite{aksan2020cose}. 

Metrics for evaluating the quality of digital ink generative models of text typically include Character Error Rate for text generation readability~\cite{chang2021style,kotani2020generating,aksan2018deepwriting}, writer identification for style transfer~\cite{kotani2020generating}, and human evaluation~\cite{kotani2020generating,aksan2018deepwriting,cao2019ai}.

Most digital ink generation approaches use either ancestral sampling or greedy sampling, with exception of~\cite{chang2022data}, which uses biased sampling~\cite{graves2013generating} for the task of generating the synthetic training data. 

To our knowledge, no studies on the effects of sampling and ranking for digital ink generation have been performed. Similarly, no studies have looked at the relationship between the generation speed and quality.

\section{Method}
\label{sec:method}

\begin{figure}
    \centering
    \includegraphics[trim=0.5cm 15.5cm 1.5cm 4.5cm,clip,width=\textwidth]{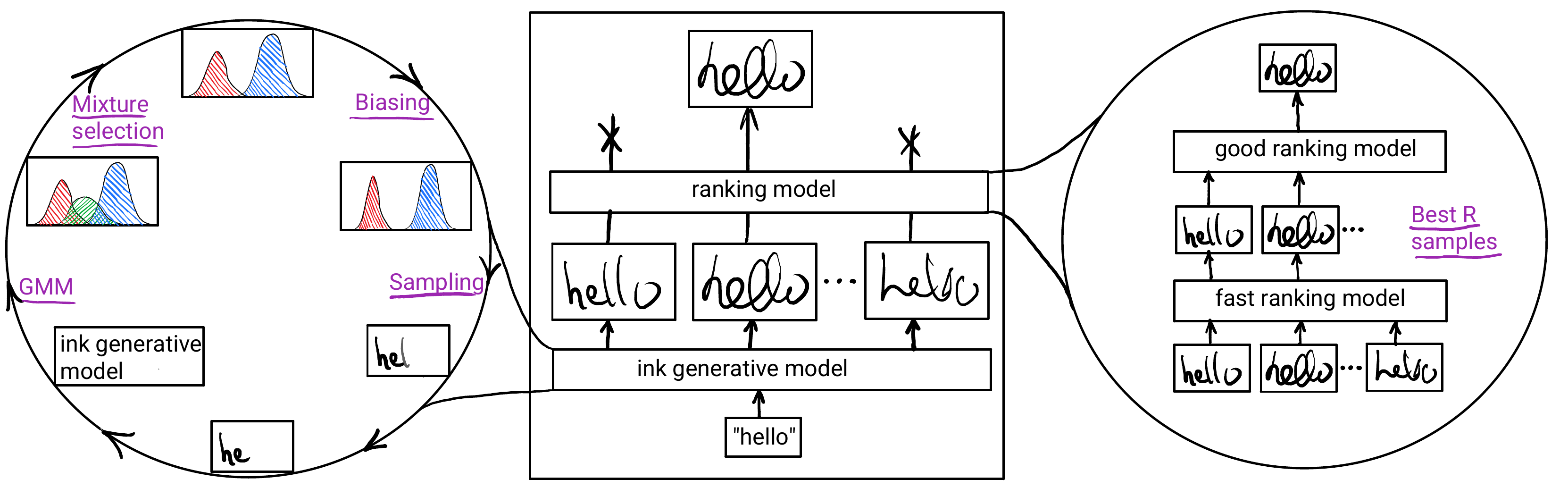}
    \caption{The diagram of the proposed solution. The input to the model is a single text label. The generative model is run to produce $B$ candidates. The highest scoring one according to the ranking model is returned. In the generative model, we use different sampling modes to modify the output distribution of the model. The ranking model consists of two blocks, first taking $B$ generated inks and scoring them, then taking the $R$ inks with the highest scores and re-ranking them.}
    \label{fig:model}
\end{figure}

Given an autoregressive generative model of digital ink that takes a text label as input and produces a sequence representing digital ink as output, we are interested in maximizing the average quality $M_{\Theta_S,\Theta_R}(S, B, R)$ of the model output, while guaranteeing that the maximum inference time does not exceed a certain threshold $\mathcal{T_{\text{max}}}$. Here, $S$ is the sampling method used by the generative model, $B$ is the size of the batch for generation, and $R$ is an inference-time parameter of the ranking model, $\Theta_S$ are fixed trained weights of the model, $\Theta_R$ are the trainable parameters of the ranking model, which we will describe below.

During inference, given a label, the generative model will use sampling method $S$ to produce a batch of $B$ digital inks, which will be scored according to the ranking model $\Theta_R$. The highest-ranking sample will be returned as the output; if $B=1$, the ranking model is bypassed. Fig.~\ref{fig:model} illustrates the approach.

Our main results concern the trade-off between the inference time and model output quality, and are presented in Sec.~\ref{sec:results}. The rest of this section is organized as follows: we describe our approach to measuring quality and inference time in Sec.~\ref{subsec:eval}; Sec.~\ref{subsec:sampling} outlines the data representation for digital ink and sampling methods $S$ that can be used with it; Sec.~\ref{subsec:ranking} describes the ranking models we use and how to train them.

\subsection{Evaluation}
\label{subsec:eval}

We propose an evaluation method linked to the system's usability.
Similar to other works~\cite{kotani2020generating,chang2022data,chang2021style,aksan2018deepwriting}, as quality measure $M$ we use the Character Error Rate (CER) of a trained handwriting recognition model on the generated samples. This stems from the assumption that the generated text is not useful if it is not readable, regardless of other attributes like style and beauty.

A second axis of interest for usability is the inference time.  We report the \textbf{worst case} inference time \textbf{per character}. We measure the worst case latency, with the assumption that exceeding the budget makes the functionality unusable for users. We measure time per character since processing time is expected to scale linearly with the sequence length. 

\subsection{Data representation and sampling}
\label{subsec:sampling}

Two frequently used representations of the digital ink data are \textbf{raw} and \textbf{curve} representation, which both encode the ink as a sequence of input tokens in $\mathbb{R}^d\times\{0,1\}^2$, with first $d$ values describing the shape of the stroke between two points, and the last 2 binary values indicating whether (i) a particular token is at the end of the stroke, and whether (ii) it is the last token in the sequence (end of ink). For the \textbf{raw} representation, $d=2$ and describes the offset between two adjacent points, and for the \textbf{curve} representation, $d=6$ and describes the parameters of Bezier curve fit to a segment of the stroke~\cite{song2020beziersketch}.

Following the approach of~\cite{graves2013generating} and most of the later literature on the topic, we parameterize the output distribution of every step of the autoregressive generative model by a set of parameters $(\pi, \mu, \Sigma, e_s, e_i)$, where $\pi, \mu, \Sigma$ describe weights, means, and covariances of a mixture of Gaussians, from which $\mathbb{R}^d$ stroke parameters are sampled, and $e_s$ and $e_i$ describe the parameters of Bernoulli distributions from which the pen-up (end-of-stroke) and end-of-sequence events are sampled. $\Sigma$ is full-covariance matrix for raw features ($d=2$) and diagonal otherwise. We provide more details in Sec.~\ref{sec:details}.

\paragraph{Sampling.} We consider two types of distortions for the output distribution:  distortion of the mixture weights $\pi$ and distortion of the diagonal components of the covariance matrix $\Sigma$. To distort the mixture weights, we consider several standard approaches: Top-K (parameterized by the value of K), and Top-P and Typical sampling (both parameterized by the value of P). To distort the covariance matrix, we subtract a \textit{sampling bias} value $b$ from the diagonal elements of the covariance matrix, before applying the softplus~\cite{glorot2011deep} function to it to ensure positive values. This reduces the variance after the model has been trained, to avoid sampling in low-confidence regions. The sampling parameters $S=(s,m,b)$ are therefore the sampling method $s\in\{\text{Top-K, Top-P, Typical}\}$, the mixture parameter $m$, and the sampling bias value $b$.  

\subsection{Ranking models}
\label{subsec:ranking}

Running a ranking model to order the generated samples may be computationally costly. For this reason, we differentiate between a process to rank all candidates and one that ranks only the most promising ones. 
Following the approach commonly used in information retrieval~\cite{liu2020asking,ravaut2022summareranker}, our ranking approach is two-staged, with a "fast" ranker $\mathcal{R}_1$ that runs on all $B$ generated outputs simultaneously, and a slower, more trustworthy "good" ranker $\mathcal{R}_2$, which is used to re-rank the samples ranked highest by $\mathcal{R}_1$. The inference time parameter $R$ of the ranking model, introduced at the beginning of this section, is the number of top samples according to $\mathcal{R}_1$ that are re-ranked by $\mathcal{R}_2$. When $R=B$, this corresponds to using only $\mathcal{R}_2$, and when $R=1$, only $\mathcal{R}_1$ is used. We describe both rankers below, and provide more details about them in Sec.~\ref{sec:setup}.

\paragraph{"Good" ranker $\mathcal{R}_2$.}  Since our goal is to generate samples with lowest possible Character Error Rate, an obvious choice for $\mathcal{R}_2$ to use the recognizer model that measures CER as the ranking model - that is, select the sample that is perfectly recognizable or has the lowest character error rate. However, running the recognizer on-device can be slow depending on the implementation, and we will see that having a faster first stage is beneficial.

\paragraph{"Fast" ranker $\mathcal{R}_1$.}
Following the approach of~\cite{TorToiSe}, our $\mathcal{R}_1$ ranker is a model learned to predict whether the generated sample is recognizable or not, that is, whether the recognizer would return the target label given the generated ink. In other words, this ranker is an approximation of the "good" ranker and tries to predict its output. Since inference time is one of the main focuses of our work, we consider a much simpler ranking model than the one described in~\cite{TorToiSe}. Instead of looking at both the generated ink and target label, our ranker just uses the generated ink. It consists of two convolutional layers followed by global average pooling. We study this choice of ranking model in terms of inference speed and the types of errors that it can address in Sec.~\ref{sec:results}.

\paragraph{Training dataset for $\mathcal{R}_1$.} As described above, $\mathcal{R}_1$ ranker is trained to be a fast approximation of the $\mathcal{R}_2$ ranker, and it predicts whether synthesized ink is even close to being recognizable. To train $\mathcal{R}_1$, we don’t use real data: we use the synthesizer for generating a sample for a given text label, and $\mathcal{R}_2$ ranker for generating a binary label of whether the sample is recognizable (recognition result matches the text label) or not. The pair of generated ink and binary label is the training data for $\mathcal{R}_1$ (more details in Sec.~\ref{sec:details}).

We first train the ranking model, and then, select the sampling method $S$ that performs best on the $\mathcal{D}_{\text{tune}}$ dataset. Doing the reverse would require training a ranking model for each possible sampling parameter setting, which would be prohibitively expensive. This means that during training of $\mathcal{R}_1$, the sampling method is yet unknown. To accommodate this, we create the training dataset for $\mathcal{R}_1$ by generating samples with $(s, m, b)$ selected at random, for each sample. This allows $\mathcal{R}_1$ to be robust to any future selection of $S$, so that the sampling parameters can be chosen after the ranker is trained. We evaluate this method of training dataset creation in Sec.~\ref{sec:results}.

\section{Results}
\label{sec:results}

\subsection{Setup}
\label{sec:setup}
To show that both sampling and ranking bring forth significant improvements in generation quality, and show the robustness of the proposed approach, we will evaluate it on 4 datasets across 3 different languages, with two frequently used model types, and two data representations.

We consider 4 digital ink datasets for text generation: English \deepwriting{}~\cite{aksan2018deepwriting} and \iamondb{}~\cite{liwicki2005iam}, Vietnamese \vnondb{}~\cite{nguyen2018database}, and an internal \maths{} dataset of mathematical expressions. We use two data representations described in Sec.~\ref{subsec:sampling}, \raw{} and \curve{}, and evaluate two different model types, \tacotron{}~\cite{graves2013generating,wang2017tacotron} and \transformer{}~\cite{vaswani2017attention}.

\subsection{Implementation details}
\label{sec:details}
For both \tacotron{} and \transformer{}, we use 10-component Gaussian mixtures in the model output. For \tacotron{}, we use one-hot encoding of labels and 3 layers of size 256 in the decoder. For \transformer{}, we use 2 layers with 4 attention heads and embedding size 64 in the label encoder, and 6 layers with 4 attention heads and embedding size 128 in the decoder. We use the Pre-LN implementation~\cite{baevski2018adaptive}. We train models with Adam with global clipnorm of 0.1, and learning rate of 1e-3 for \tacotron{} and learning rate schedule described in~\cite{vaswani2017attention} for \transformer{}. Models are trained for $2\times 10^6$ steps with batch size 256. For training the $\mathcal{R}_1$ ranker, we generate $10^5$ samples with labels from the generator training data as the training set, and 1000 samples with labels from the generator validation data as the validation set. As described in Sec.~\ref{sec:method}, for each sample, we select a sampling method at random to generate it. The pool of sampling methods includes $\text{Top-P, Typical}$ samplings with $m\in\{0.0,0.1,\dots,1.0\}$ and $\text{Top-K}$ sampling with $m\in\{1,2,\dots,10\}$, and sampling biases $b\in\{0,1,5,25,100,\infty\}$. The $\mathcal{R}_2$ ranker is a state-of-the-art recognizer that has been trained on internal data not related to public datasets and is an LSTM-CTC model with 6 layers of size 216~\cite{carbune}, which is combined with word and character language models during beam search decoding, similar to~\cite{keysers2016multi}.

For \iamondb{}, we use \textit{testset\_v} for validation, \textit{testset\_f} for tuning sampling parameters (via grid search over all possible samplings), and \textit{testset\_t} for testing. For \vnondb{}, we use the version of the dataset split by individual words. Since this dataset does not have the tuning subset, we use validation data labels for tuning sampling parameters. For \deepwriting{}, since this dataset does not have tuning or testing subset, we extracted 1500 labels whose lengths have the same mean and variance as the \deepwriting{} validation data, from the labels present in the \textbf{IAMonDO} dataset (we include these labels with the submission for clarity). Models were implemented in Tensorflow and the time measurements were done after conversion to TFLite on a Samsung Galaxy Tab S7+ tablet.

\subsection{Baselines}

\paragraph{Sampling model baseline.} We compare the model with tuned sampling parameters, with a model with fixed sampling method. Since different works in the literature consider different sampling methods, to have a fair comparison to them, as to a baseline, we report the best result with $S=(\text{Top-P},m,b), m\in\{0.0, 1.0\}, b\in\{0.0,\infty\}$, that is, greedy or ancestral sampling of component with infinite or zero bias for the offset parameters. We will refer to the optimal sampling method as $S_{\text{opt}}$, and to baseline as $S_{\text{base}}$.

\paragraph{Ranking model baseline.} We compare the $\mathcal{R}_1$ ranker that predicts the recognizability of the generated ink, described in Sec.~\ref{sec:method}, with an approach described in~\cite{RankGen}, which trains a model to distinguish between real and synthesized samples, with the goal of selecting the most "real-looking" samples. We will refer to it as $\mathcal{R}_{\text{base}}$.

\subsection{Quantitative analysis}

\begin{table*}
\small
\centering
\caption{CER for different sampling and ranking strategies. For $S_{\text{base}}$ and $S_{\text{opt}}$, we use $B=1$, meaning that no ranker is used. For $\mathcal{R}_1$ and $\mathcal{R}_{\text{base}}$, we use $B=5$ and $R=1$, meaning that "good" $\mathcal{R}_2$ ranker is not used. For $\mathcal{R}_2$, we use $B=5$ and $R=5$, meaning that the samples are ranked according to the "good" ranker only. This number is also a bound on the quality achievable with a "fast" ranker $\mathcal{R}_1$. }
\begin{tabular}{lllrrrrr}
\toprule
\textbf{Dataset}  &
\textbf{Data} &
\textbf{Model}  &
$S_{\text{base}}$ &
$S_{\text{opt}}$ &
$\mathcal{R}_{\text{base}}$ &
$\mathcal{R}_1$ &
$\mathcal{R}_2$ \\
\midrule

\multirow{4}{*}{\deepwriting} & \multirow{2}{*}{\raw} 
                             & \tacotron& 4.6\tiny{$\pm0.6$} & 2.6\tiny{$\pm0.2$} & 2.3\tiny{$\pm0.3$} & 1.7\tiny{$\pm0.2$} & 0.7\tiny{$\pm0.1$} \\
                             &          & \transformer& 8.1\tiny{$\pm2.9$} & 6.7\tiny{$\pm1.8$} & 5.8\tiny{$\pm1.3$}& 4.9\tiny{$\pm1.1$} & 1.8\tiny{$\pm0.5$} \\
                             & \multirow{2}{*}{\curve} & \tacotron& 5.9\tiny{$\pm0.5$} & 5.6\tiny{$\pm0.7$} & 4.5\tiny{$\pm 0.7$} & 2.1\tiny{$\pm0.2$} & 0.9\tiny{$\pm0.1$} \\
                             &          & \transformer& 8.9\tiny{$\pm1.5$} & 6.6\tiny{$\pm0.9$} & 4.7\tiny{$\pm0.5$} & 2.8\tiny{$\pm0.3$} & 1.0\tiny{$\pm0.1$} \\
\hline
\multirow{4}{*}{\iamondb} & \multirow{2}{*}{\raw} & \tacotron& 5.8\tiny{$\pm3.1$} & 3.8\tiny{$\pm0.7$} & 3.7\tiny{$\pm0.9$} & 2.6\tiny{$\pm0.4$} & 1.3\tiny{$\pm0.1$} \\
                             &          & \transformer& 13.3\tiny{$\pm2.9$} & 12.3\tiny{$\pm2.0$} & 10.9\tiny{$\pm0.2$} & 9.3\tiny{$\pm1.2$} & 5.3\tiny{$\pm1.2$} \\
                             & \multirow{2}{*}{\curve} & \tacotron& 14.9\tiny{$\pm1.2$} & 9.1\tiny{$\pm0.9$} & 9.1\tiny{$\pm0.6$} & 3.8\tiny{$\pm0.0$} & 2.1\tiny{$\pm0.1$} \\
                             &          & \transformer& 16.8\tiny{$\pm1.4$} & 12.0\tiny{$\pm1.6$} & 11.7\tiny{$\pm1.0$} & 8.2\tiny{$\pm0.4$} & 3.9\tiny{$\pm0.7$} \\
\hline
\multirow{4}{*}{\vnondb} & \multirow{2}{*}{\raw} & \tacotron& 4.0\tiny{$\pm0.5$} & 3.2\tiny{$\pm0.6$} & 3.2\tiny{$\pm0.5$} & 2.1\tiny{$\pm0.2$} & 0.7\tiny{$\pm0.1$} \\
                             &          & \transformer& 4.3\tiny{$\pm0.9$} & 3.7\tiny{$\pm0.6$} & 3.0\tiny{$\pm0.4$} & 2.6\tiny{$\pm0.4$} & 0.8\tiny{$\pm0.1$} \\
                             & \multirow{2}{*}{\curve} & \tacotron& 2.1\tiny{$\pm0.1$} & 2.2\tiny{$\pm0.2$} & 2.2\tiny{$\pm0.2$} & 1.8\tiny{$\pm0.2$} & 0.7\tiny{$\pm0.1$} \\
                             &          & \transformer& 2.0\tiny{$\pm0.2$} & 2.0\tiny{$\pm0.2$} & 2.0\tiny{$\pm0.2$} & 1.8\tiny{$\pm0.3$} & 0.7\tiny{$\pm0.0$} \\
\hline
\multirow{4}{*}{\maths} & \multirow{2}{*}{\raw} & \tacotron& 28.5\tiny{$\pm1.0$} & 23.1\tiny{$\pm1.1$} & 22.3\tiny{$\pm0.4$} & 18.5\tiny{$\pm0.6$} & 8.3\tiny{$\pm0.5$} \\
                             &          & \transformer& 28.1\tiny{$\pm4.0$} & 22.8\tiny{$\pm2.5$} & 20.3\tiny{$\pm3.0$} & 19.7\tiny{$\pm2.9$} & 8.3\tiny{$\pm1.1$} \\
                             & \multirow{2}{*}{\curve} & \tacotron& 9.4\tiny{$\pm0.5$} & 9.4\tiny{$\pm0.6$} & 9.0\tiny{$\pm0.1$} & 9.0\tiny{$\pm0.1$} & 3.1\tiny{$\pm0.1$} \\
                             &          & \transformer& 13.6\tiny{$\pm1.8$} & 10.8\tiny{$\pm0.7$} & 9.6\tiny{$\pm0.6$} & 9.2\tiny{$\pm0.4$} & 4.0\tiny{$\pm0.1$} \\

\bottomrule
\end{tabular}
\label{tab:comparison}
\end{table*}

\paragraph{Effect of sampling and ranking}

In Table~\ref{tab:comparison}, we compare the results of applying different sampling and ranking techniques for all datasets, model types, and data types.  

A first major finding of our study is that \textbf{tuning the sampling technique helps in almost all cases} - in 13 cases out of 16, with the remaining ones being ties.

The second conclusion is that using a ranking model helps \textbf{in all cases}. 

There is still a significant gap between the performance when using $\mathcal{R}_1$ and the quality-optimal $\mathcal{R}_2$. However, as we show in the next paragraph, achieving such quality comes with penalties for inference time.

Finally, we can conclude that using ranker that predicts whether the ink is recognizable or not is superior to using a baseline ranker~\cite{RankGen} that predicts whether a given ink is real or synthetic. However the latter ranker also helps in most cases, as compared to not using ranking at all.

\paragraph{Comparison under a time budget.}
\label{par:time}

\begin{figure}[ht!]
    \centering
\includegraphics[width=0.9\linewidth]{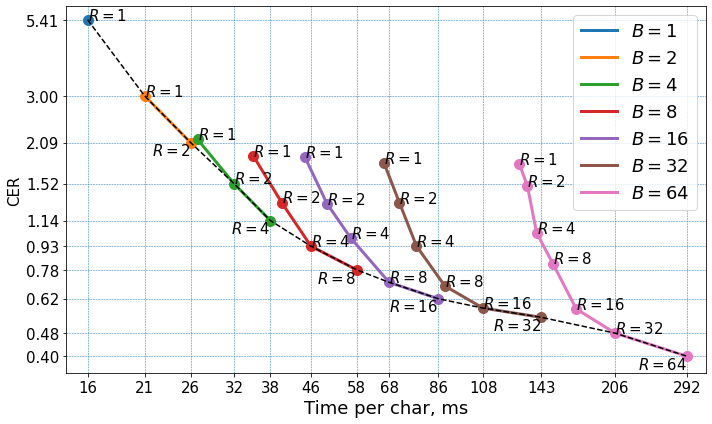}
\caption{Model inference time (upper bound, per char) vs CER for various values of $B$ and $R$. For each values of $B$, we report results for values of $R$ in $\{1, 2, 4, 8, 16, 32, 64\}$ s.t. $R\le B$. The gray dotted line shows a Pareto-optimal frontier. Both axes on the log-scale. As visible, there are points on the Pareto frontier that include the use of both $\mathcal{R}_1$ and $\mathcal{R}_2$, justifying our claim that there are scenarios where optimal performance for a given computational budget can be achieved by a combination of both.}% \cmusat{This is important -  the  time per char is the time per batch}}
\label{fig:batching}
\end{figure}

The inference time for the model consists of 3 separate parts: \textit{(i)} generating a batch of $B$ samples; \textit{(ii)} ranking them with the $\mathcal{R}_1$ ranker (unless $B=R$, in which case we can use just $\mathcal{R}_2$); \textit{(iii)} Re-ranking the top $R$ candidates with $\mathcal{R}_2$ (unless $B=1$ in which case the generated sample can be returned directly). We show how these values scale with the input batch size for the model (that is, $B$ for generative model and $\mathcal{R}_1$, and $R$ for $\mathcal{R}_2$), in Table~\ref{tab:batching}, and the trade-off between CER and inference time in Fig.~\ref{fig:batching}.

\begin{table}[ht!]
\centering
\caption{Model inference time per character, in milliseconds, for generative model, ranking model $\mathcal{R}_1$, and recognizer $\mathcal{R}_2$. Average across 1000 labels, \tacotron{} model on \deepwriting{} data with \curve{} data representation. The generation process can be efficiently vectorized and scales sub-linearly. The inference time of $\mathcal{R}_1$ is almost negligible, and the inference time of $\mathcal{R}_2$ scales linearly.}

\begin{tabular}{llrr}
\toprule
\textbf{Batch size}  & \textbf{Generation} & \textbf{$\mathcal{R}_1$} & \textbf{$\mathcal{R}_2$} \\

1 & 15.5 & 0.05 & 2.79 \\
2 & 20.6 & 0.05 & 5.19 \\
4 & 26.6 & 0.09 & 11.40 \\
8 & 35.0 & 0.15 & 23.04 \\
16 & 45.0 & 0.24 & 41.39 \\
32 & 66.3 & 0.45 & 76.97 \\
64 & 128.6 & 0.91 & 163.47 \\

\bottomrule
\end{tabular}
\label{tab:batching}
\end{table}

Here we present the comparison of model quality vs inference time budget, by varying the values of $B$ and $R$.

To connect the input sequence length to inference time, we fix the maximum number of decoding steps the model is allowed to make per input sequence symbol. In other words, our inference time is measured as time needed for one decoding step times the maximum allowed number of tokens per input symbol. The generation is always run until the maximum number of frames. In the models we used for this evaluation, 99\% of the samples generated less than 5 frames per output character, which is the ratio that we fixed.

Table~\ref{tab:batching} shows the inference time for synthesis model, $\mathcal{R}_1$, and $\mathcal{R}_2$, in ms per character as a function of the input batch size. Notice that both the autoregressive generative model and the convolution-based ranker are able to take advantage of vectorization and are 7.5 and 3.2 times faster for large batch sizes than if run individually. The recognizer, used as $\mathcal{R}_2$, however, does not parallelize well due to CTC~\cite{graves2006connectionist} decoding and combination with language models, thus scaling linearly with the batch size.

Based on the data in Table~\ref{tab:batching}, we plot the numbers for model quality and worst-case inference time for different values of $B$ and $R$ in Fig.~\ref{fig:batching}. Points with $(B=4,R=2)$, $(B=8,R=4)$, and $(B=16,R=8)$ are on the Pareto frontier, verifying our earlier statement that there are scenarios where the best performance can be achieved by combining the two rankers. Points $(B=2,R=1)$ and $(B=4,R=1)$ are also on the frontier, verifying our statement that there are cases where the best performance can be achieved without using the recognizer part of the ranking model at all.

\paragraph{Discussion and limitations.} We note that the findings we present here are not universal, and the exact inference time depends on a multitude of factors such as specific generative model type and size, hardware, length of the sequence to be generated (processor caching makes longer sequences faster on a per-character basis), ranking model type and size (for the recognizer ranker, we rely on a model using CTC decoding which is hard to vectorize, whereas Seq2Seq models may parallelize better, although usually have worse accuracy). Furthermore, the average/median inference time might differ from the worst case significantly: The generative model produces an average 3.7 output frames per input character, compared to 5 which we used for the worst case analysis. Also when using the recognizer as a ranker, we need not recognize all of the candidates as we can stop at the first candidate that is perfectly recognizable, which may happen sooner or later depending on the exact sampling type and model quality. However, we believe that this does not invalidate our findings: depending on the time budget, better performance may be achieved by using a fast learned ranking model or combining it with a recognizer.

\paragraph{Ablation study.}
\begin{table*}
\small
\centering
\caption{Ablation study for the ranker. The first column contains the results obtained when using $\mathcal{R}_1$ as the ranker. The next group of columns ablates the way of constructing the training dataset - by always generating samples using ancestral sampling, or by always generating samples using the sampling that yields the optimal performance when using $\mathcal{R}_2$ as the ranker. The last column shows that the optimal sampling parameters are different for each setup, ablating our choice of always tuning the sampling parameters.}
\begin{tabular}{ccc|c|cc|l}
\toprule
\multirow{2}{*}{\textbf{Dataset}}  &
\multirow{2}{*}{\textbf{Data}} &
\multirow{2}{*}{\textbf{Model}} & 
\multirow{2}{*}{\textbf{$\mathcal{R}_1$}} &
\multicolumn{2}{c|}{\textbf{Ranker training data}} &
\multirow{2}{*}{\textbf{Opt. sampling}}  \\
& & & &  \multicolumn{1}{c}{Anc.} & \multicolumn{1}{c|}{Rec.} &
\\

\midrule

\multirow{4}{*}{\deepwriting} & \multirow{2}{*}{\raw} 
& \tacotron &  1.7\tiny{$\pm0.2$} &  1.9\tiny{$\pm0.2$} & 2.0\tiny{$\pm0.2$} & Top-P, 0.9, 5.0 \\ 
& & \transformer & 4.9\tiny{$\pm1.1$} & 5.4\tiny{$\pm1.0$} & 5.0\tiny{$\pm0.9$} & Top-K, 9, $\infty$  \\
& \multirow{2}{*}{\curve} 
& \tacotron& 2.1\tiny{$\pm0.2$} &  2.0\tiny{$\pm0.4$} & 2.0\tiny{$\pm0.4$} & Top-K, 3, $\infty$ \\ 
& & \transformer& 2.8\tiny{$\pm0.3$} & 2.7\tiny{$\pm0.3$} & 2.8\tiny{$\pm0.3$} & Top-K, 5, $\infty$  \\
\hline
\multirow{4}{*}{\iamondb} & \multirow{2}{*}{\raw} & 
\tacotron & 2.6\tiny{$\pm0.4$} & 2.8\tiny{$\pm0.5$} & 2.6\tiny{$\pm0.4$} & Top-P, 0.9, 100.0 \\ 
& & \transformer & 9.3\tiny{$\pm1.2$} & 9.1\tiny{$\pm1.3$} & 9.3\tiny{$\pm1.5$} & Top-K, 6, $\infty$  \\ 
& \multirow{2}{*}{\curve} &
\tacotron & 3.8\tiny{$\pm0.0$} & 3.8\tiny{$\pm0.1$} & 4.3\tiny{$\pm0.3$} & Top-K, 2, $\infty$  \\
& & \transformer & 8.2\tiny{$\pm0.4$} & 8.6\tiny{$\pm0.8$} & 8.2\tiny{$\pm0.8$} & Top-K, 4, $\infty$  \\  
\hline
\multirow{4}{*}{\vnondb} & \multirow{2}{*}{\raw} &
\tacotron & 2.1\tiny{$\pm0.2$} & 2.5\tiny{$\pm0.2$} & 2.4\tiny{$\pm0.2$} & Top-P, 0.9, 100.0 \\
& & \transformer & 2.6\tiny{$\pm0.4$} & 2.8\tiny{$\pm0.4$} & 2.9\tiny{$\pm0.4$} & Top-P, 0.9, 5.0  \\
& \multirow{2}{*}{\curve} 
& \tacotron & 1.8\tiny{$\pm0.2$} & 2.0\tiny{$\pm0.1$} & 1.7\tiny{$\pm0.1$} & Top-P, 0.4, $\infty$ \\
& & \transformer & 1.8\tiny{$\pm0.3$} & 2.8\tiny{$\pm0.4$} & 2.9\tiny{$\pm0.4$} & Top-P, 0.3, $\infty$  \\
\hline
\multirow{4}{*}{\maths} & \multirow{2}{*}{\raw} 
& \tacotron & 18.5\tiny{$\pm0.6$} & 19.4\tiny{$\pm0.6$} & 19.0\tiny{$\pm0.6$} & Top-P, 0.9, 5.0  \\ 
& & \transformer & 19.7\tiny{$\pm2.9$} & 20.5\tiny{$\pm2.7$} & 20.0\tiny{$\pm2.1$} & Top-K, 8, $\infty$  \\
& \multirow{2}{*}{\curve} &
\tacotron & 7.7\tiny{$\pm0.3$} & 8.4\tiny{$\pm0.1$} & 7.7\tiny{$\pm0.2$} & Top-P, 0.3, $\infty$ \\
& & \transformer & 9.2\tiny{$\pm0.4$} & 10.2\tiny{$\pm0.5$} & 9.3\tiny{$\pm0.1$} & Top-P, 0.3, $\infty$  \\

\bottomrule
\end{tabular}
\label{tab:ablation}
\end{table*}

In Table~\ref{tab:ablation} we evaluate our choice of the construction of the ranker training dataset, and tuning of the sampling parameters for every setup (generation model type and feature type). 

Firstly, we compare our approach of generating training data for the ranker by using random sampling parameters for every label to two other baseline approaches: \textit{(i)} using a fixed ancestral sampling when generating the training data; this intuitively makes sense as sampling from "widest" possible distribution should cover all the whole diversity of the generated data. \textit{(ii)} for each setup, using the sampling parameters that yield the lowest CER if $\mathcal{R}_2$ is used  as ranker; this makes sense as $\mathcal{R}_1$ tries to approximate $\mathcal{R}_2$, and it is reasonable to assume that their optimal sampling parameters should be similar. We observe that on average our proposed way of constructing a training dataset is optimal, never being more than one decimal point worse than other approaches, but at times significantly outperforming them.

Secondly, we show that the optimal sampling parameters differ a lot between the setups, so it is important to tune them for each setup. The only reliable signals we observed was that for the \curve{} representation, it is often preferable to sample more "greedily" (lower value of K in Top-K or P in Top-P sampling) than for the \raw{} representation, and that the optimal samplings seem to be somewhat close between the two model types.

\subsection{Qualitative analysis}

\begin{table}[ht!]
\centering
\caption{Number of overconfidence and incoherence errors for various values of $p$ in Top-P sampling, for a model with and without $\mathcal{R}_1$ ranker.} 
\begin{tabular}{l|cc|cc}
\toprule
\multirow{2}{*}{\textbf{P}}  & \multicolumn{2}{c}{\textbf{No ranking}} & \multicolumn{2}{c}{\textbf{Ranking with $\mathcal{R}_1$}} \\
& Overconf. & Incoher. & Overconf. & Incoher. \\
\midrule
0.1 & 81 & 120 & 42 & 157  \\
0.2 & 75 & 115 & 37 & 114 \\
0.3 & 69 & 140 & 23 & 111 \\
0.4 & 59 & 170 & 16 & 109 \\
0.5 & 41 & 180 & 9 & 109 \\
0.6 & 33 & 216 & 3 & 121 \\
0.7 & 30  & 246 & 2 & 137 \\
0.8 & 22 & 281 & 1 & 149 \\
0.9 & 14 & 375 & 1 & 197  \\
1.0 & 7 & 466 & 1 & 282 \\
\bottomrule
\end{tabular}
\label{tab:errors}
\end{table}

\begin{figure}[ht!]
\centering
\vspace{-1cm}
\includegraphics[width=0.9\linewidth]{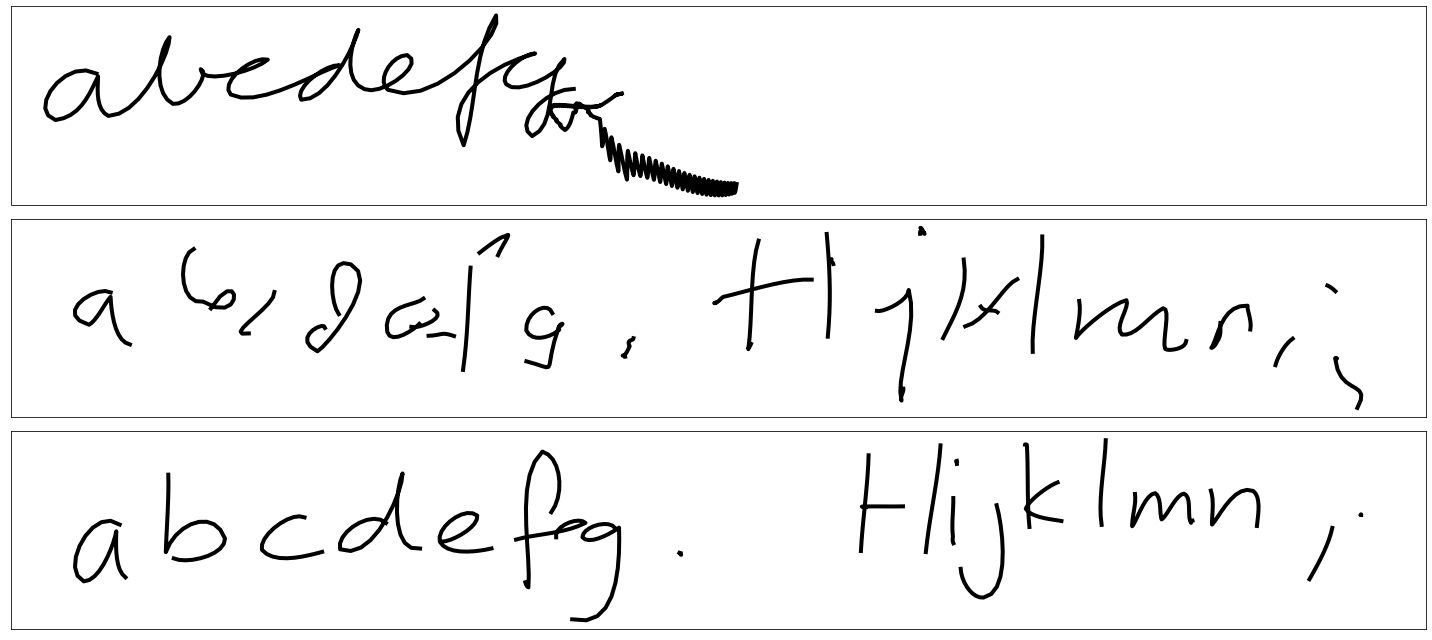}
\captionof{figure}{Examples of model outputs for different sampling parameters. Input label is "\textit{abcdefg. Hijklmn,}". Sampling parameters used are: Top - (Top-P, 0.0, $\infty$); Middle - (Top-P, 1.0, 0.0); Bottom - (Top-P, 0.5, 5.0). The overconfidence error is clearly visible in the top example, while the middle example is incoherent and hard to recognize. The bottom row shows the importance of carefully selecting sampling for optimal performance.}
\label{fig:errors}
\end{figure}

\begin{figure*}
    \centering
    \includegraphics[width=\textwidth]{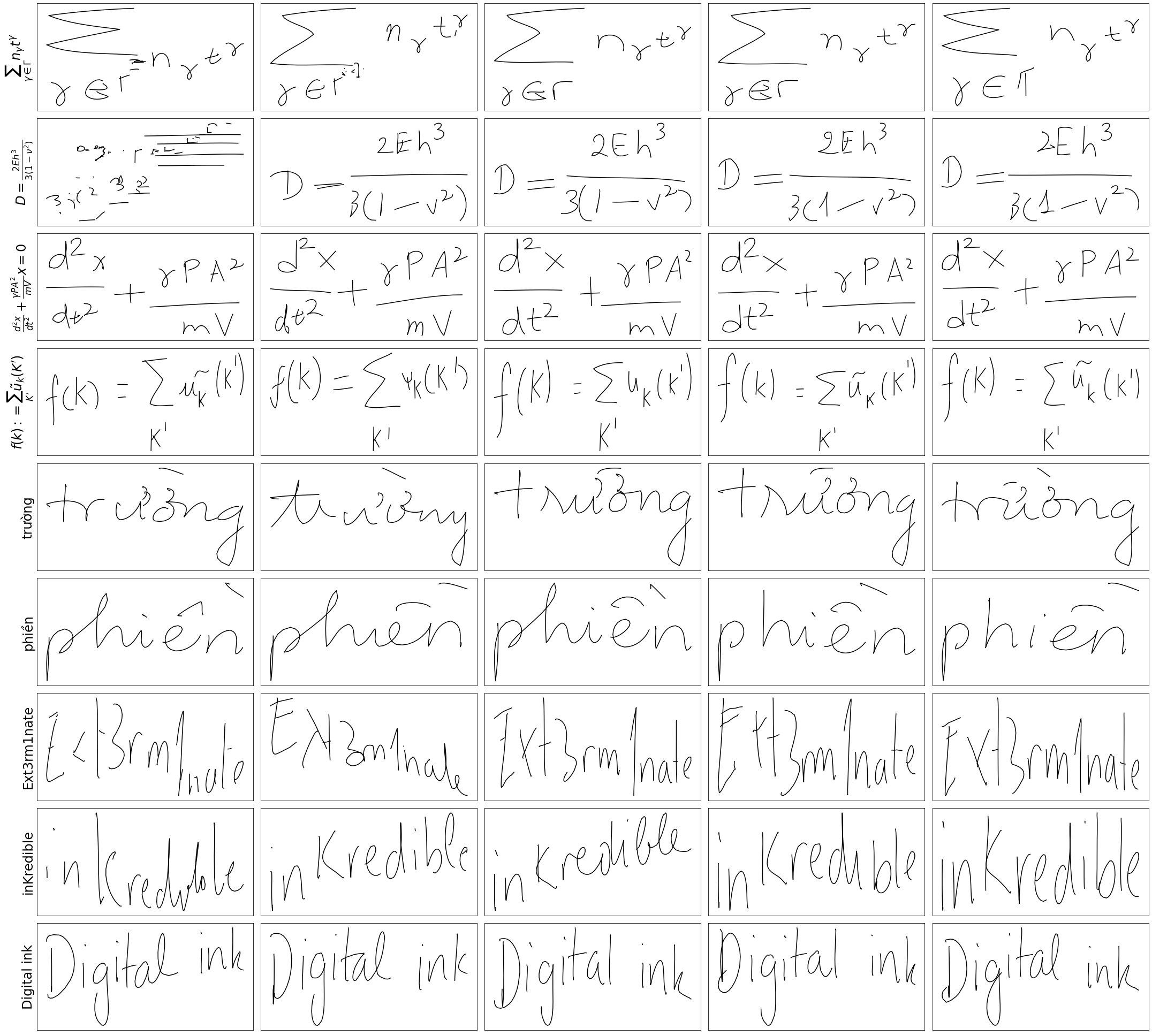}
    \caption{Examples of model outputs. \transformer{} with \curve{} representation for \maths{} data, \tacotron{} with \curve{} representation for \vnondb{} data, \tacotron{} with \raw{} representation for \deepwriting{} data. In each case, 5 samples were generated, and sorted left-to-right according to the score provided by the $\mathcal{R}_1$ ranker model, with the rightmost image being the most recognizable according to the ranker. The first column shows some examples of samples that are not recognizable and are scored low by the ranker, ex. stray strokes (first row), overconfident generation of repeated lines (second row), misplaced tilde sign over u (fourth row), one extra diacritic (fifth row), missing dash over t (seventh row). More examples can be obtained in demo colab: \url{https://colab.research.google.com/drive/1AkwmDOkEIkifbOYEBdcB9PrR_Ll-fcmz}.}

    \label{fig:examples}
\end{figure*}

In this section, we first attempt to confirm that: \textit{(i)} the two types of errors, overconfidence and incoherence, actually happen when generating digital ink samples, and \textit{(ii)} both the choice of sampling and ranking has effect on these errors. Results are presented with the \textbf{Tacotron} model on \textbf{Deepwriting} dataset with \textbf{curve} representation, but we have observed largely similar trends for other cases. Afterwards, we present examples of model output on various datasets.

Fig~\ref{fig:errors} shows examples of generated ink with various samplings - with both incoherence and overconfidence examples visible. As we can observe, overconfidence errors typically result in very long ink, that can not be recognized as the label, with repeating pattern inside. Given this observation, we attempt to quantify the number of errors of each type by looking at \textit{samples that can not be recognized} (meaning the label returned by the recognizer differs from the input label to the generative model), and within those samples, whether the generation process reached the maximum number of steps (implying overconfidence) or not (implying incoherence). Table~\ref{tab:errors} shows the number of errors, estimated by this approach, as a function of sampling parameters (value of p in Top-P sampling), and it confirms the intuition about how it should behave. We can see that as the sampling parameters go from greedy sampling closer to ancestral sampling, the number of overconfidence errors goes down, while the number of incoherence errors goes up. When we use the ranking model, we see that the number of incoherence samples first goes down, and then goes up. We attribute this to the fact that as sampling becomes more diverse, the ranking model is able to select better candidates, but as sampling becomes too diverse, all candidates start being less recognizable. Overall, using ranking seems to reduce the number of overconfidence errors by 50-90\%, and number of incoherence errors by up to 50\%.

Fig.~\ref{fig:examples} shows of the model outputs, sorted according to the score provided by the ranker, left-to-right. As can be seen, the rightmost sample in every row is recognizable and matches the label, while the leftmost sample is mostly not recognizable. It is expected that in many cases at least one of 5 samples is not recognizable - if that were not the case, that would mean that the selected sampling method is too conservative and should be relaxed to produce samples with higher diversity (which would trade-off having all 5 candidates recognizable in "easy" cases for improved performance in "difficult" cases where all 5 samples were not recognizable).

\section{Conclusion}
In this paper, we investigated the effects of combining sampling and ranking strategies to improve digital ink generation.

These methods, used before in other domains such as NLG and TTS, proved to be highly useful, and complementary to each other in the case of digital ink. Until now, however, they were not explored in this domain, with most methods using ancestral or greedy sampling, and no candidate ranking.
We evaluate sampling and ranking techniques, on four datasets - two containing writing in English and one in Vietnamese, as well as a fourth one with mathematical formulas. We test the robustness of the findings using two model types (Tacotron and Transformer) and two common ink data representations (\raw{} and \curve{}). In all the combinations, we report significant improvements in the recognizability of the synthetic inks: taken together, a well-chosen sampling method, followed by fast ranking consistently improve recognizability, in many cases halving the character error rates.

An important factor in the perceived quality of ink synthesis is speed. Potential applications, such as handwriting autocompletion, spelling correction, and beautification usually process user inputs on-device, so ink generative models need to be fast. We thus report the findings with respect to a given computational budget.

\bibliographystyle{splncs04}
\bibliography{ijcai22}

\end{document}